\documentstyle[preprint,aps]{revtex}

\begin{document}
\draft
\title{Ohmic Losses in Valence-band Photoemission Experiments}
\author{Robert Haslinger and Robert Joynt}
\address{Department of Physics and Applied Superconductivity Center\\
University of Wisconsin-Madison \\
1150 University Avenue \\
Madison, WI 53706 \\
}
\date{\today}
\maketitle

\begin{abstract}
Photoemission experiments involve the motion of an electron near a
conducting surface. This necessarily generates heat by ohmic losses from
eddy currents.  This inelastic scattering of the electrons will result in a downward
shift in observed spectra. This effect is
most pronounced in poorly conducting metals: in good metals the electron's
field is screened out of the material, while insulators are by definition unable to
absorb electromagnetic energy at low frequencies.  We give a classification of 
photoemission processes which shows that the effect is an extrinsic
process distinct from final state effects.  The shift is illustrated by a model
system with a Drude-like conductivity function and a temperature-dependent
relaxation time.  We give a brief experimental survey of systems in which 
the ohmic losses may be significant.
\end{abstract}

\newpage

\section{Introduction}

Photoemission provides the most direct way to measure the the energy
spectrum of electrons in a solid. The equation given in elementary 
modern physics texts in
the section on the photoelectric effect is 
\begin{equation}
E = E_{\nu} + \epsilon - W,  \label{eq:pe}
\end{equation}
where $E$ is the energy of the emitted electron, $E_{\nu}$ is the energy of
the incident photon, $\epsilon$ is the energy of the electron in the solid
relative to the chemical potential and W is the work function. For
monochromatic incident light, the spectrum of observed energies $E$ is the
same as that for the electron energies $\epsilon$ in the solid, except for a
known shift.

This equation is known to be oversimplified in a number of respects. For example,
an electron can scatter from other electrons on leaving the solid, giving up
some of its energy, and even possibly producing secondary electrons which
can escape from the sample. In fact, the mean free path of an electron moving
with kinetic energies of 20 eV is short, of order $10 \AA$, while the
penetration depth of the light is much longer, meaning that only a small
fraction of the electromagnetic energy absorbed actually leads to a
photoemitted electron satisfying Eq.\ \ref{eq:pe}. Nonetheless, Eq.\ \ref
{eq:pe} is used for the analysis of the vast majority of angle-integrated
photoemission spectroscopy (AIPES) experiments. The justification for this
is the belief that the electrons that absorb a photon fall into two categories:
(1) electrons that scatter strongly inelastically, losing an energy of several
electron volts, and (2) electrons that do not scatter at all.
Electrons in the first category either do not escape from the solid, or they
do so with such a low kinetic energy that they can be discarded as
uninteresting. Those in the second category are believed to follow Eq.\ \ref
{eq:pe}. In view of the short mean free path, they must come from a thin
layer near the surface, leading to the characterization of photoemission
(PE) as a surface-sensitive probe.

Once stated in this way, a possible difficulty with this
reasoning suggests itself.  The argument depends on having a 
clean separation of energy scales.  If an electron loses an intermediate
amount of energy, the logic breaks down.  In
modern experiments, spectral features on a scale of 10 - 100 meV are
very often of great interest. If many electrons have a 
substantial probability of
undergoing inelastic scatttering with that much energy loss, the use of Eq.\ 
\ref{eq:pe}, which forms the basis of the interpretation of nearly
all PE data, is no longer correct.

This paper will argue that this separation cannot always be be made,
and that the assumptions underlying Eq.\ \ref{eq:pe} can be dangerous in
materials with relatively low conductivity or low effective carrier
density.  This includes many systems of
current experimental interest. 

Consider the photoemitted electron after it
has left the solid at a speed $v$. Its electric field creates eddy currents
and ohmic losses in the solid it is leaving behind. These losses are
determined by the dielectric function $\epsilon(\omega)$. If the spectral
features are of width $\hbar \omega \approx$ 10 - 100 meV, then it is the
Drude (or at least the low-frequency) dielectric response that determines the 
extent of the low-energy inelastic
scattering. Dimensional considerations to be detailed 
below show that the probability for low energy scattering is 
\begin{equation}
P_1 = \frac{e^2}{v \hbar} {\cal A}.
\label{eq:est}
\end{equation}
Here $e^2/\hbar v = (1/137) (c/v) \approx 0.8$ for $E_{\nu} = mv^2/2 = 20$ eV.
${\cal A}$ is a
dimensionless number that measures the absorption strength of the material
in the Drude regime, but it is generally of order unity for
isotropic materials with a Drude-like response.  Thus the fraction of
electrons that scatter is not small.  The important issue is the typical energy
loss of the electron.  We will estimate below that a typical loss is
of order $\hbar \sqrt{\sigma_0/\tau}$ 
where $\sigma_0$ is the DC conductivity and $1/\tau$ is a relaxation rate -
the width of the Drude portion of the function $\sigma(\omega)$. This is
in the important 10-100 meV regime for materials in which the effective carrier density
is low.  This condition is fulfilled in many materials of current interest;
in particular a material which undergoes a metal-insulator 
transition will usually pass through this regime.

Loss processes of the kind treated here, but generally at higher energy, 
have been considered
before. For example, plasmon emission in X-ray PE causes satellite peaks in
the measurement of core-level energies \cite{xray}.  Inelastic losses in
EELS (electron energy loss spectroscopy) are due to essentially the same physical
process \cite{eels}.
The theoretical description of these experiments is
formally similar to that given here.  The difference lies in the 
application of the concepts.  The Drude part of the dielectric function
is often temperature-dependent.  This dependence can be dramatic in 
certain cases, as, for example, at a superconducting
transition or a metal-insulator transition.  This means that the extrinsic losses
are also temperature-dependent.  It is then not permissible to interpret 
temperature-dependent changes in an AIPES spectrum as being due only to 
changes in the density of states (DOS).  The Drude part may also be 
sample-dependent.  Thus, changes in AIPES from sample to sample may reflect changes 
in the conductivity, as well as changes in the DOS.  Such effects are not 
so likely to occur in the plasmon part of the dielectric function.
The low-frequency end of these theories has, however, never been of great 
interest before for PE.  It is only the great advances in PE
resolution, now $\leq 20$ meV for many experiments, which lends new interest
to the application of these known concepts.  A short summary of the results has
appeared in previous publications \cite{me}.

In this paper we first give a classification of PE processes in order to situate the 
process treated here in the context of the 
overall theory of PE.  Next, a detailed derivation of the loss 
function is presented, and its effect on observed spectra is illustrated in a simple 
case.  Finally we give a short survey of some experimental systems that are 
candidates for the effect.

\section{Classification of Loss Processes}

The appropriate formalism is that due to 
Schaich and Ashcroft \cite{ashcroft} and Caroli {\it et al.} \cite{caroli}.  
We present here a graphical interpretation of this formalism.  

The conductivity (absorption of a low energy photon) in 
a solid is described by a current-current 
correlation.  In the language of Feynman diagrams, we have a line representing an 
electron and a line representing a hole.  Together, these make a 
loop with two vertices, the dipole matrix elements, 
at the ends.  The PE experiment also involves the absorption of a photon, but
there are two important differences between a conductivity diagram and a PE diagram
for angle-resolved photoemission (ARPES).
First, there is a measurement of the energy and momentum of the 
electron.  Second, the electron line corresponds to a vacuum state, not a bulk state.

We represent the measurement by a rectangle inserted in the electron line.  
(In order to give formal expressions corresponding to each diagram, it is necessary
to use the Keldysh rather than the Feynman formulation of perturbation theory,
but the distinction does not affect the qualitative considerations given here.)
This may now be called a current-current-current correlation function, as the 
electron line has been split into two.  (The nomenclature is somewhat misleading 
since it would seem to imply that PE measures a third-rank tensor. It does not.)  
      
The simplest diagram for a PE  process is shown in Fig.\ 
\ref{fig:fig1}.  The electron propagates freely to the detector in this case.  
The measurement fixes the energy and momentum of the 
electron line, so it 
contributes only a delta function to the 
diagram and what we have is the textbook process 
of Eq.\ \ref{eq:pe}.  If the momentum is averaged over, then the
experiment is AIPES, which is the experiment 
we will concentrate on in this paper.

Fig.\ \ref{fig:fig2} is the most-studied of all the 
diagrams.  The thick solid line is the full propagator for the hole.
This diagram represents intrinsic many-body effects.  All many-body effects on the 
propagation of a hole in the bulk system must be included, in principle.  The reason 
for calling this diagram "intrinsic", is that nothing about 
the photoemission process itself is involved.  The emitted electron still 
propagates freely to the detector.  The diagram of Fig.\ \ref{fig:fig2}
includes that of Fig.\ \ref{fig:fig1}.  To the extent 
that only these diagrams are important, PE measures the spectral function of a 
hole.  In simpler language, it measures the occupied DOS.  

Fig.\ \ref{fig:fig3}
is the effect of the interaction of the emitted electron with the hole left behind.  
This is usually called a final state interaction, an "extrinsic" process, as it 
depends on the PE process itself.  The rate of such interactions 
decreases with incoming photon energy, and this diagram is usually neglected.  This 
is called the sudden approximation.  Fig.\ \ref{fig:fig4} represents an extrinsic 
process in the Born approximation.  The thick interaction line is fully screened, and 
thus includes interactions with all the electrons in the material.  This changes 
the energy and the momentum of the observed particle.  This diagram is the 
subject of the present work.  Note that the distinction between the final state 
interactions and the extrinsic inelastic loss processes is a clean one.  The two 
should not be confused.  The ohmic losses are, roughly speaking, due to interactions 
with the dynamic image charge, not the photohole.

This is not a complete listing of all possible processes, as one can also 
have scattering from impurities of the photoelectron, reflection at the surface, 
and so on.  We have tried to give a reasonable overview of the possibilities that 
make an interesting comparison or contrast with the ohmic process.
 
\section{Derivation for Cubic Systems}

We begin with the general problem of a sample that occupies the half-space $z<0$.  
There is a time-dependent charge density $\rho(\vec{r},t)$ outside it.
The Fourier component at frequency $\omega$, $\rho(\vec{r},\omega)$, sets 
up an electric field for $z<0$ given by
\begin{equation}
\vec{E}(\vec{r}, \omega) = \int_{z'>0}
\frac{\rho(\vec{r}', \omega)}{|\vec{r}-\vec{r}'|}
\frac{d^3 r'}{1 + \epsilon(\omega)}.
\label{eq:field}
\end{equation}
This result follows from the usual image-charge 
calculation, namely the solution of
Maxwell's equations with the standard surface boundary conditions.  Nevertheless, 
there are certain approximations.  The 
normal skin effect is assumed because the wavevector dependence of $\epsilon$ is 
neglected.  This could break down for clean materials or at quite high 
frequencies.  Fortunately, both of these conditions are outside of the 
applications which interest us here.  Secondly, the real surface is not of 
infinitesimal thickness, and corrections for this have been neglected.  Finally, 
we have specialized to the case that the material 
has cubic symmetry: the electric field and the electric displacement 
at any given frequency are 
simply proportional.
The field drives currents that 
generate heat.  The total energy thus produced is:
\begin{equation}
Q = \frac{1}{2}  \int_{z<0} d \omega \Re \sigma(\omega)
|\vec{E}(\vec{r}, \omega)|^2 d^3 r.
\label{eq:heat}
\end{equation}

These results are quite general for any charge density
for which the integral in Eq.\ \ref{eq:heat} is convergent.
Now we specialize to the PE process, which is modeled  
as a charge generated at the surface at 
time $t=0$ leaving the solid at 
an angle $\theta$ to the normal.  The charge density is
\begin{equation}
\rho(\vec{r},t) = - e \Theta(t) \delta(z - v t \cos \theta) \delta(y) 
\delta(x - vt \sin \theta), 
\end{equation}
where $\Theta(t)$ is the step function.  
Since in most experiments $\theta \leq \pi/10$, we set $\cos \theta = 1$
henceforth.  The Fourier transform is
\begin{equation}
\rho(\vec{r}, \omega) = \frac{-e}{2 \pi v \cos \theta} 
\delta(x) \delta(y) e^{-i \omega z /v} \Theta(z).
\end{equation}  
Using Eq.\ \ref{eq:field}, we have
\begin{equation}
\vec{E}(\vec{r}, \omega) =
\frac{-e}{2 \pi v \cos \theta} \frac{2}{1 + \epsilon(\omega)}
\int_0^{\infty} d z' e^{-i \omega z'/v \cos \theta}
\frac{\vec{r} - z' \hat{z}}{|\vec{r} - z' \hat{z}|^3}
\end{equation}
as the field inside the sample.
Substituting this into Eq.\ \ref{eq:heat} yields
\begin{equation}
Q = \frac{2 e^2}{\pi v} {\cal C}
\int_0^{\infty} d \omega \frac{\Re \sigma(\omega) }{\omega |1 + \epsilon 
(\omega)|^2}, 
\end{equation}
where ${\cal C}$ is a dimensionless integral:
\begin{equation}
{\cal C} = \int_0^{\infty} \rho d \rho \int_0^{\infty} dz
\left( \rho^2 |I_1(\rho, z)|^2 + |I_2(\rho, z)|^2 \right),
\end{equation}
with
\begin{equation}
I_1 = \int_0^{\infty} \frac{e^{-i \zeta} d \zeta}{\left[\rho^2+(z + \zeta)^2 
\right]^{3/2}}
\end{equation}
and 
\begin{equation}
I_2 = \int_0^{\infty} \frac{e^{-i \zeta} (z + \zeta) d \zeta}
{\left[\rho^2+(z + \zeta)^2 
\right]^{3/2}}.
\end{equation}
${\cal C}$ may be estimated as about 2.6.  

This classical calculation yields not only the total energy
loss but also the loss into each frequency interval $ d \omega$.
Quantum-mechanically, this is interpreted as the inelastic energy loss
at this energy due to a scattering event from the bulk electrons.
It was calculated assuming no change in the trajectory of the electron,
and is therefore essentially the Born approximation.  Fig.\ \ref{fig:fig4} 
could also 
have several scattering lines.  This would correspond to approximations
beyond Born.  When such processes are important, then the amount
information contained in
a PE spectrum decreases, just as in the case of multiphonon
events in neutron scattering (for example).  

The relative differential probability to lose 
energy $\hbar \omega$ is 
\begin{equation}
P(\omega) = \frac{2 e^2 {\cal C}}{\pi \hbar v \omega^2} 
\frac{\Re \sigma (\omega)}{ |1+\epsilon(\omega)|^2}.
\label{eq:po}
\end{equation}

There is also a probability for
forward scattering $P_0$.  This is the probability that an electron
loses zero energy.  
Including this possibility, the observed intensity is given by 
\begin{equation}
I(\omega, T) = P_0(T) N(\omega) f(\omega) 
+ \int_{- \infty}^{\infty} P(\omega-\omega', T) N(\omega') f(\omega') d \omega',
\label{eq:int}
\end{equation}
Here 
$N(\omega)$ is the DOS and $f$ is the Fermi function.
The total normalization is given by our assumption that there
is either one or zero scatterings:
\begin{equation}
1 = P_0 + \int_0^{\infty} P(\omega) d \omega.
\end{equation}
Equations having the form of Eq.\ \ref{eq:int}
are well-known in considerations of background in PE.
For example, the Shirley background would result from the phenomenological assumption 
that $P(\omega)$ is a constant to be fit to experiment \cite{shirley}.

\section{Experimental consequences}

>From the point of view of experiment, one would like 
rules of thumb for recognizing the situations in which the ohmic
effect is likely to be important.  To this end, we investigate 
more closely the function $P(\omega)$.  Note first that 
\begin{equation}
P(\omega) = \frac{2 e^2 {\cal C}}{\pi \hbar v} 
\frac{\Re \sigma (\omega)}{\omega^2 |1 + \epsilon(\omega)|^2}
\end{equation}
vanishes at small frequencies if the system is
a true insulator, since $\Re \sigma$ vanishes faster than
any power of $\omega$.  This must be true from a physical
standpoint: an insulator cannot absorb electromagnetic energy.
If the system is a very good metal,
then $ \sigma_0 \equiv \Re \sigma(\omega=0)$ is large, and    
at low frequencies 
$\epsilon(\omega) \rightarrow 4 \pi i \sigma_0/\omega$
leading to
\begin{equation}
P(\omega) \rightarrow
\frac{1}{16 \pi^2 \sigma_0} \frac{2 e^2 {\cal C}}{\pi \hbar v}.
\end{equation}
The vanishing of $P(\omega)$ in this limit is also required by elementary 
physics.  A perfect metal must completely screen out the 
field at sufficiently low frequencies.  
We conclude that materials in the intermediate range are of most interest.
Certainly if PE is done across a metal-insulator transition
this range is explored.  

There is no general formula that determines the 
importance of ohmic losses, because $\epsilon(\omega)= 
1 + 4 \pi i \sigma / \omega$ can take many forms, with the consequence that
even materials with similar DC conductivities may give quite different
results.  Furthermore, the effect depends on the frequency range which 
is being explored.  However, it very often 
happens that in the low-frequency regime
the two-parameter Drude function  
\begin{equation}
\sigma(\omega) = \frac{\sigma_0}{1 - i \omega \tau}
\label{eq:dr}
\end{equation}
is a reasonable approximation.  
In this paper, we will focus on this particular form.  Even for materials
for which the Drude form is not a very good quantitative
approximation for the conductivity, it may still be true that one parameter
for the overall scale and one for the width are sufficient.  
In this case, the results below should give a reasonable guide.

The Drude form for $P(\omega)$,
obtained by substituting Eq.\ \ref{eq:dr} into Eq.\ \ref{eq:po}
and rearranging, is:
\begin{equation}
P(\omega)= \frac{2 e^2 {\cal C}}{\pi \hbar v}
\frac{\sigma_0\tau^2}{4}
\left\{\left[\omega^2 \tau^2 - (2 \pi \sigma \tau-1/2)\right]^2 + 
2 \pi \sigma_0 \tau - 1/4 \right\}^{-1}. 
\label{eq:pint}
\end{equation}

The first thing to note about this function is its normalization.
Integrating Eq.\ \ref{eq:pint} we find 
\begin{equation}
\int_0^{\infty}P(\omega) d \omega = \frac{e^2 {\cal C}}{ 4\pi \hbar v}.
\end{equation}
In 
terms of the notation of Eq.\ \ref{eq:est}, this says ${\cal A} = {\cal C}/4
\approx 0.65$.  Substituting in a velocity corresponding to the 
energy $E_{\nu}=20$ eV we find $P_1 = 0.5$.
Hence the integrated effect is substantial.  Since this result is
independent of the parameters, it is a generic result 
not limited to the particular details of the 
Drude form.  Bear in mind that this is not the
only contribution to $P_1$ since losses at higher energies, for
example losses due to interband processes, will also contribute.
On the other hand, the fact that $P_1$ is comparable to unity at 
typical experimental energies also implies that we near the limit 
of validity of the Born approximation.

Written in terms of the variable $\omega^2 \tau^2$, $P(\omega)$ is a Lorentzian
centered at $2 \pi \sigma \tau-1/2 $ and width 
$\sqrt{2 \pi \sigma_0 \tau - 1/4}$.  In terms of the physical variable
$\omega$, we have two regimes.  If $2 \pi \sigma_0 \tau < 1/2$, then the 
function is peaked at $\omega = 0$, and falls off at large frequencies
as $\omega^{-4}$ on a characteristic scale of $1/\tau$. 
If $2 \pi \sigma_0 \tau > 1/2$, then the 
function is peaked at 
\begin{equation}
\omega_{peak} = \sqrt{2 \pi \sigma_0 / \tau - 1/ 2 \tau^2} \rightarrow 
\sqrt{2 \pi \sigma_0 / \tau }, 
\end{equation}
and the width of the peak is 
\begin{equation}
\Delta \omega = \frac{1}{\tau} 
\left(2 \pi \sigma_0 \tau - 1/4 \right)^{1/4}
\rightarrow \left(\frac{2 \pi \sigma_0}{ \tau^3} \right)^{1/4}.
\end{equation}
The limiting behaviors for $2 \pi \sigma_0 \tau >>1$ are also given. 

The overall behavior of $P(\omega)$ is illustrated in Fig.\ \ref{fig:pomega}.
It is plotted for a fixed value of $\sigma_0/\tau$.  In the Drude model
\begin{equation}
\frac{\sigma_0}{\tau} = \frac{ne^2}{m^*} = \frac{\omega_p^2}{4 \pi},
\end{equation}
where $\omega_p$ is an effective plasma frequency.  Thus, in a system
where the effective carrier concentration is fixed and the 
temperature dependence comes either from the relaxation
time or from $\sigma_0$, Fig.\ \ref{fig:pomega} represents the evolution of
$P(\omega)$ with temperature.  This graph is for a low effective carrier
density: $\sigma_0 / \tau = 10^4$ meV$^2$, which corresponds to
an effective plasma frequency of 354 meV.  The relaxation times are in
units of meV$^{-1}$ where $\hbar=1$, that is $6.6 \times 10^{-13}s$.
As $\tau$ increases, the peak position moves from $\omega=0$
to the limiting value
$\sqrt{2 \pi \sigma_0 / \tau }$, while it continually sharpens up.

These structures in $P(\omega)$ manifest them selves in the
observed intensity according to Eq.\ \ref{eq:int}.
Consider the case of zero temperature.  Then $f(\omega)$ becomes
$\theta(-\omega)$ and we have:
\begin{equation}
I(\omega)= N(\omega)\theta(-\omega)P_0 + 
\int_{\omega}^{0} P(\omega'-\omega)N(\omega') d \omega'.
\end{equation}
We are only interested in the second term, the inelastic component, here.
Furthermore, we shall take a constant DOS $N(\omega)=N_0=1$ so that the 
features in the observed intensity come only from the energy dependence of the\
scattering.  Thus we normalize $P(\omega)$ to unity.
\begin{equation}
I_{inelastic}(\omega)= \int_{\omega}^{0} P(\omega'-\omega) d\omega'
\end{equation}
The integral rearranges the apparent DOS.  The DOS will
appear unchanged at frequency $\omega$ if $|\omega| >>
\omega_{peak}$ where $\omega_{peak}=0$ or 
$\sqrt{\frac{2 \pi \sigma_0}{\tau} -\frac{1}{2\tau^2}}$
depending on the size of $\sigma_0 \tau$.  The photoelectrons observed
at large $|\omega|$ are shifted, but it {\it appears as if}
the (flat) DOS is unaffected at these frequencies.
At low frequencies the situation is more complicated.
The basic scenario for a constant $\sigma_0 / \tau$ (effective carrier
density) is this:
at 'small' $\tau<<\tau_c$ 
$P(\omega)$
is peaked around $\omega=0$ so the PE loss is negligible.
At the crossover lifetime 
$\tau_c= 
\sqrt{\frac{m^*}{4\pi n e^2}}$
the peak of $P(\omega)$ begins to shift to a higher frequency.
In addition, $P(\omega)$ is very spread out for $\tau$ in this
region, so the apparent loss becomes large and may appear pseudogap.
When $\tau>>\tau_c$ the shape of the 
observed intensity $I(\omega)$ looks unchanged,
but the entire curve is shifted downwards by an amount comparable to
the effective plasma frequency, and this may look like a gap.
This evolution of $I(\omega)$ with increasing $\tau$ is shown 
in Fig.\ \ref{fig:intensity}.  Bear in mind that this is only 
the inelastic component and it must be combined with the 
elastic intrinsic part.  This may involve fitting since the
overall normalization of the elastic part $(P_0)$ would normally be
hard to determine. 
  
It should be noted that while most of the {\bf peak}
shift of $P(\omega)$ occurs over a relatively small range of $\tau$ the evolution
of $I(\omega)$ takes place over several orders of magnitude of $\tau$ since
$P(\omega)$ is spread out over a large frequency range.
We can see this by plotting the shift of the intensity curve
vs. $\tau$ as in Fig.\ \ref{fig:width}.  
The shift is defined by the position in energy of the point where
the intensity is half its maximum, a popular definition of the gap
in PE experiments.
 
\section{Survey of Experimental Systems}

Experimentally, the best candidates for gap-like structures are the
one-dimensional conductors.  There are several 
one-dimensional materials that show the 
development of some sort of a gap in the spectrum without
corresponding observations in other properties.  
We mention here only two.  

TTF-TCNQ is a one-dimensional organic chain compound.  It has a $2k_F$
charge-density-wave transition at $T_P = 54 K$ with an energy gap $2 
\Delta \sim 40$ meV, as measured by the activation energy of transport and
thermodynamic quantities.  The 
photoemission spectrum \cite{zwick} is very much at odds with the transport 
and thermodynamics.  At $70 K$, there is a gap of about $0.15 \sim 0.2 eV$, 
judging by the peak in an energy scan at an angle that
corresponds to $k_F$.  This peak does 
not move toward the Fermi energy as T increases.  Rather, the spectrum 
just broadens out, but without increasing the weight at the Fermi 
energy.  In fact, the weight if anything decreases \cite{grioni}.
This material is a good candidate for a pseudogap of partly 
extrinsic origin.

(TaSe$_4$)$_2$I presents a somewhat similar scenario.  It has a $2 k_F$ 
charge density wave
transition at about $250 K$.  A number of its 
properties: resistivity, magnetic susceptibility, and optical 
conductivity can be explained by assuming that the gap is roughly 250 
meV, and remains at about this value, independent of temperature, even 
after long-range positional order is lost \cite{nic}.  But the gap in 
photoemission is about 500 meV, independent of temperature 
\cite{dardel}.  Here we 
may have a gap which is split about 50-50 between intrinsic and 
extrinsic.  

The theory presented above for a Drude-like conductivity
is not applicable to the one-dimensional case as there is certainly a genuine 
pseudogap or gap, depending on temperature, and the systems are not cubic.  
The factor of two between the PE gap and the
gap inferred from other properties in (TaSe$_4$)$_2$  We
would find a natural explanation
if the extrinsic loss turns on at the intrinsic gap frequency.      

The colossal magnetoresistance manganate materials are good candidates for extrinsic
losses in the neighborhood of their metal-insulator transition.  The 
cubic material La$_{0.67}$Ca$_{0.33}$MnO$_3$ is in fact the only system for which 
model calculations of the extrinsic losses have been carried out 
\cite{me}.  The material has a
metal-insulator transition at $260~K$.  
In the 'metallic' state at $80~K$, there is a strong 
negative slope in $I(\omega)$ for at least 0.6 eV below $\mu$ 
\cite{park}.  
There is a sharp break in slope at $\mu$, 
presumably indicative of a nonzero
density of states at $\mu$.  But in the insulating state at $280~K$
there appears to be no
Fermi edge at all - the observed intensity is flat at $\mu$
and weight has moved back from $\mu$. 
These features might be taken as indicating the presence of a pseudogap 
which which opens in the insulating state.  However, this movement of spectral weight 
can be produced by extrinsic effects, as shown in \cite{me}.

The layered material La$_{1.2}$Sr$_{1.8}$Mn$_2$O$_7$, on the other hand, has 
been shown by Dessau and Saitoh \cite{ds} to have a momentum-dependent pseudogap 
somewhat similar to the high-T$_c$ materials, but rather larger, in the 
range of $200 meV$ or so at the maximum.  Again, the momentum dependence 
indicates that the pseudogap is mostly of intrinsic origin. 
The layered manganate mterials are an interesting case because of their 
extremely high resistivities.  La$_{1.2}$Sr$_{1.8}$Mn$_2$O$_7$ itself has an 
a-b plane resistivity of about 6 m $\Omega$-cm at T = 50 K, which is in the 
'metallic' phase.  This is an extraordinarily high value.  It is so close to
being insulating that one would probably not expect ohmic effects \cite{me2}.  

Another class of poorly conducting materals are the Kondo insulators.  
This class of systems shows a resistivity dominated by the Kondo effect 
at higher temeratures crossing over to activated behavior at low 
temperatures, with absolute values which can be quite high.  The 
resistivity as a function of temperature is monotonic, in contrast to the 
CMR systems.  For example, the system YbB$_{12}$ has such a crossover 
at about $50 K$.  This material also shows some indications of extrinsic 
processes.  Even on the metallic side, the observed density of states 
slopes downward strongly to the Fermi energy \cite{yb}.  As the resistivity 
increases, there is a further depression of weight near $E_F$, with 
finally a gap-like structure of about 20 meV appearing at 6 K.

The best-known pseudogap of all is that in the high-temperature 
superconductor Bi$_2$Sr$_2$CaCu$_2$O$_8$ \cite{loeser}.  In this case, the 
pseudogap is both momentum- and temperature-dependent, appearing mainly 
along the Cu-O bond direction.  This is a strong indication that the 
pseudogap is largely intrinsic, as the extrinsic losses should not have 
a strong momentum dependence.  
Bi$_2$Sr$_2$CaCu$_2$O$_8$ is roughly tetragonal and the theory developed for
the cubic case is not applicable.  Its conductivity in the 
a-b plane is fairly high.  The c-axis conductivity, which is 
very low, also plays a role, however.  The theory for
such a strongly anisotropic system has not been worked
out in detail, and the combined effect of $\epsilon_{xx}$ and 
$\epsilon_{zz}$ is not yet completely clear.  it seems most likely
that lineshapes are affected by extrinsic 
effects, but that the size of the pseudogap is correctly given 
by the naive analysis.  

The high-temperature superconductors and related cuprates
are a class of systems for which
some energy-dependent PE has been done.  This is very interesting from the
point of ohmic effects, which are predicted to decline as $1/v \sim 1/\sqrt{E_{\nu}}$,
as seen from Eq.\ \ref{eq:po}.
For example, Sr$_2$CuO$_2$Cl$_2$ has been 
investigated using ARPES for $E_{\nu}=22 eV$ and $E_{\nu}=35 eV$ \cite{haffner} and
Pb-doped Bi$_2$Sr$_2$CaCu$_2$O$_8$ at $E_{\nu}=32 eV$, $E_{\nu}=40 eV$ and
$E_{\nu}=50 eV$ \cite{legner}.  In every case, changes in intensity as a function
of $E_{\nu}$ are seen.  In the case of 
Pb-doped Bi$_2$Sr$_2$CaCu$_2$O$_8$ they go in the direction
of increasing binding energy as $E_{\nu}$ is decreased, which is in
qualitative agreement with the present scenario.
In addition, the lineshapes are rather broad and also energy dependent.

It is evident from this brief survey that an important direction for
development of the theory would be a generalization to tetragonal
and orthorhombic materials, as relatively few of the most interesting
systems are cubic.  The image calculation can be performed analytically
for the tetragonal case if the emission is along the
c-axis, but the loss function is rather complicated
and will be published elsewhere.  The orthorhombic case
is even more complicated, as the image method is not applicable.   

\section{Conclusion}

Changes in the density of states of electronic systems on the scale
of 10 to 100 meV are of great interest in today's PE experiments.
These energies often coincide with the scale of relaxation rates
and conductivities or combinations thereof. 
This means that ohmic losses may affect the interpretation
of the data.  This effect is distinct from others, such as final state
effects, that have been considered in detail in the past.

General formulas for the loss in terms of the dielectric function
may be given.  In the case where a Drude form is appropriate,
we may characterize the situation as follows.
The inelastic loss will be most
pronounced if the lifetime is roughly $\tau_c \approx 1/(2 \pi \sigma_0)$.
The magnitude of the loss is on the scale of the maximum
peak shift 
$\omega_{peak}= \sqrt{2\pi \sigma_0 / \tau}$ near the effective plasma frequency.
If the effective plasma frequency is large, as in a good metal,
then the inelastically scattered electrons are emitted well below
the chemical potential as secondaries, and the elastic peak is unsullied.
If the product $\sigma_0 \tau$ is very small, as in a near-insulator,
the scattered electrons are emitted with very small loss
and cannot be distinguished from the unscattered ones.
In the intermediate case, however, a combination of the elastic and
inelastic terms is required to give the observed spectrum.
This ohmic effect may account for the observation of a gap or pseudogap
in some systems with relatively low intrinsic conductivities
in one or more directions.  In general, however, low conductivity is a necessary,
not a sufficient, condition, for the ohmic effects to be important.  
The effective carrier density and the frequency range of the measurements
must also be taken into account.

\section{Acknowledgements}
We would like to thank M. Rzchowski, M.B. Webb, A.V. Chubukov,
L. Bruch, M. Norman, J.`Joyce, and F.J. Himpsel for helpful 
discussions.  This work was supported by the NSF Materials Theory 
Program, Grant No. DMR-0081039, and by the Materials Research Science 
and Engineering Center Program, Grant No. DMR-96-32527.

\begin{figure}
\caption[]{The simplest diagram for the calculation of photoemission
intensities.  A photon (wavy line) of energy E$_{\nu}$ is absorbed by the system,
creating an electron in a vacuum state (right-going line) with energy
$E$ and a non-interacting hole in a bulk state (left-going line) with energy $\epsilon$.  
The measurement of the electron energy
and momentum is represented by the dark rectangle, and distinguishes this
diagram from an ordinary conductivity diagram.  If this diagram
is the only important one, the measured intensity in AIPES can be interpreted as 
the non-interacting DOS. }
\label{fig:fig1}
\end{figure}

\begin{figure}
\caption[]{Modification of the simplest diagram to include bulk interactions
of the hole, indicated by the thick line.  This diagram includes the previous one.  
If this diagram is the only important one, the measured intensity in AIPES 
can be interpreted as the interacting DOS.}
\label{fig:fig2}
\end{figure}

\begin{figure}
\caption[]{The diagram for final state effects in which the hole interacts with
the outgoing electron.}
\label{fig:fig3}
\end{figure}

\begin{figure}
\caption[]{The diagram for scattering of the outgoing electron 
from other electrons in the material.  This includes the extrinsic
ohmic process.}
\label{fig:fig4}
\end{figure}

\begin{figure}
\caption[]{Evolution of the relative differential probability $P(\omega)$
for losing an energy of $\hbar \omega$.  For the different plots, the effective
carrier density is held fixed: $\sigma_0 / \tau = 10^4$ meV$^2$, while
the relaxation times are in
units of $meV^{-1}$ where $\hbar=1$, that is $6.6 \times 10^{-13}s$.}
\label{fig:pomega}
\end{figure}

\begin{figure}
\caption[]{The observed {\it inelastic} intensity corresponding to the
loss functions in Fig. 5.}
\label{fig:intensity}
\end{figure}

\begin{figure}
\caption[]{The shift of the inelastic intensity curve
vs. $\tau$ as in Fig.\ \ref{fig:width}.  
The shift is defined by the position in energy of the point where
the intensity is half its maximum.}
\label{fig:width}
\end{figure}

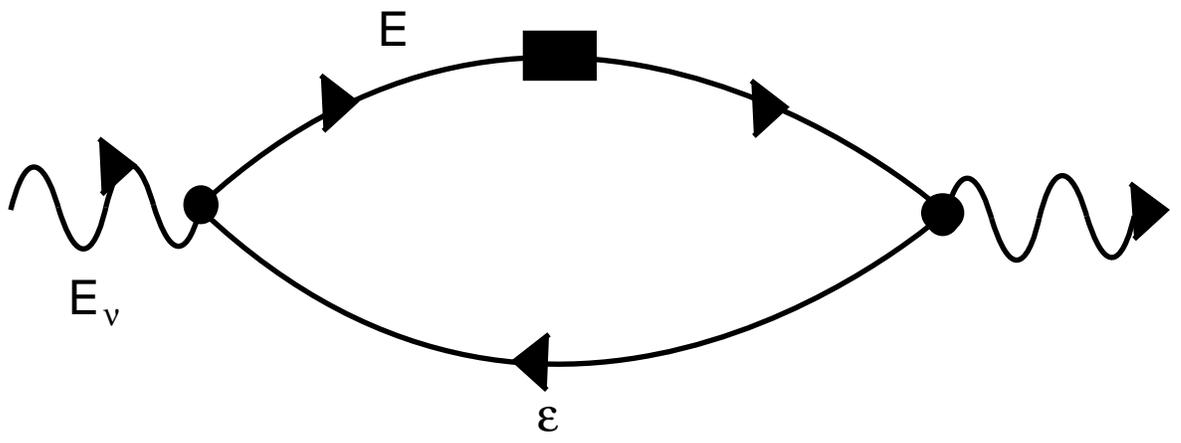
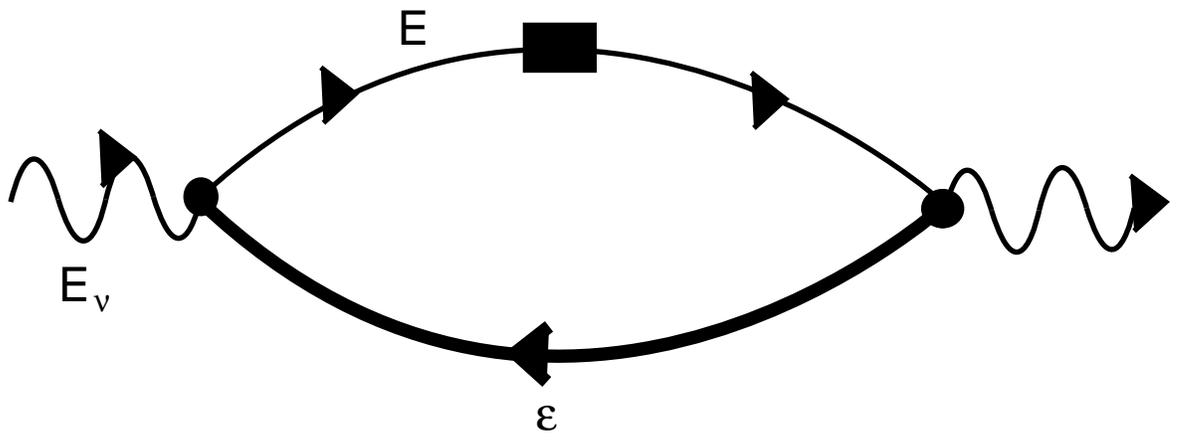
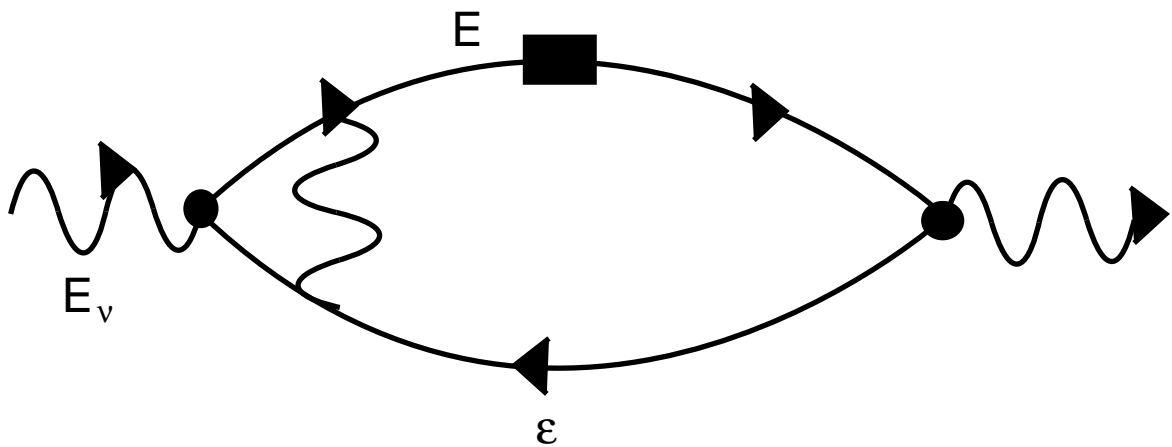
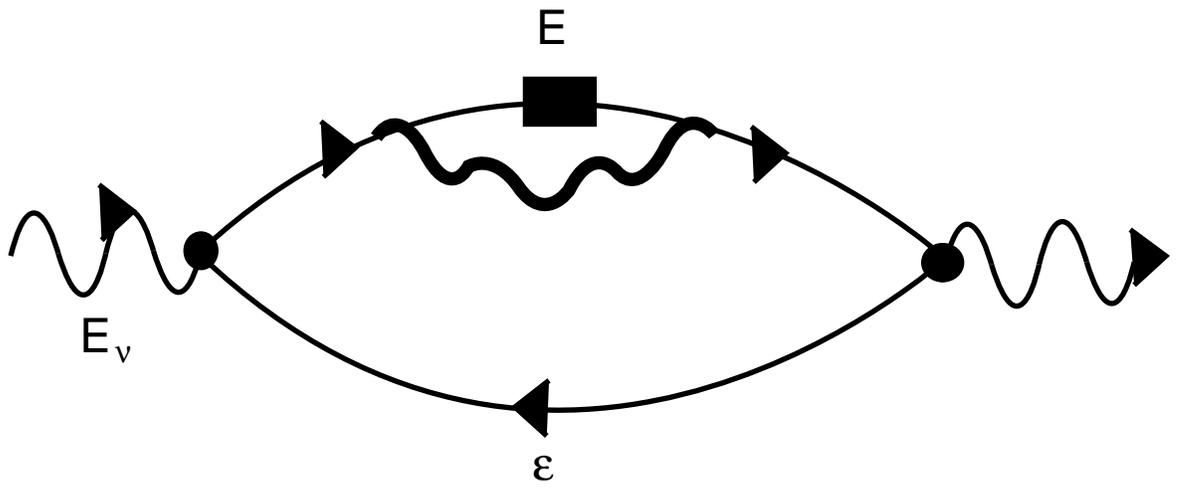
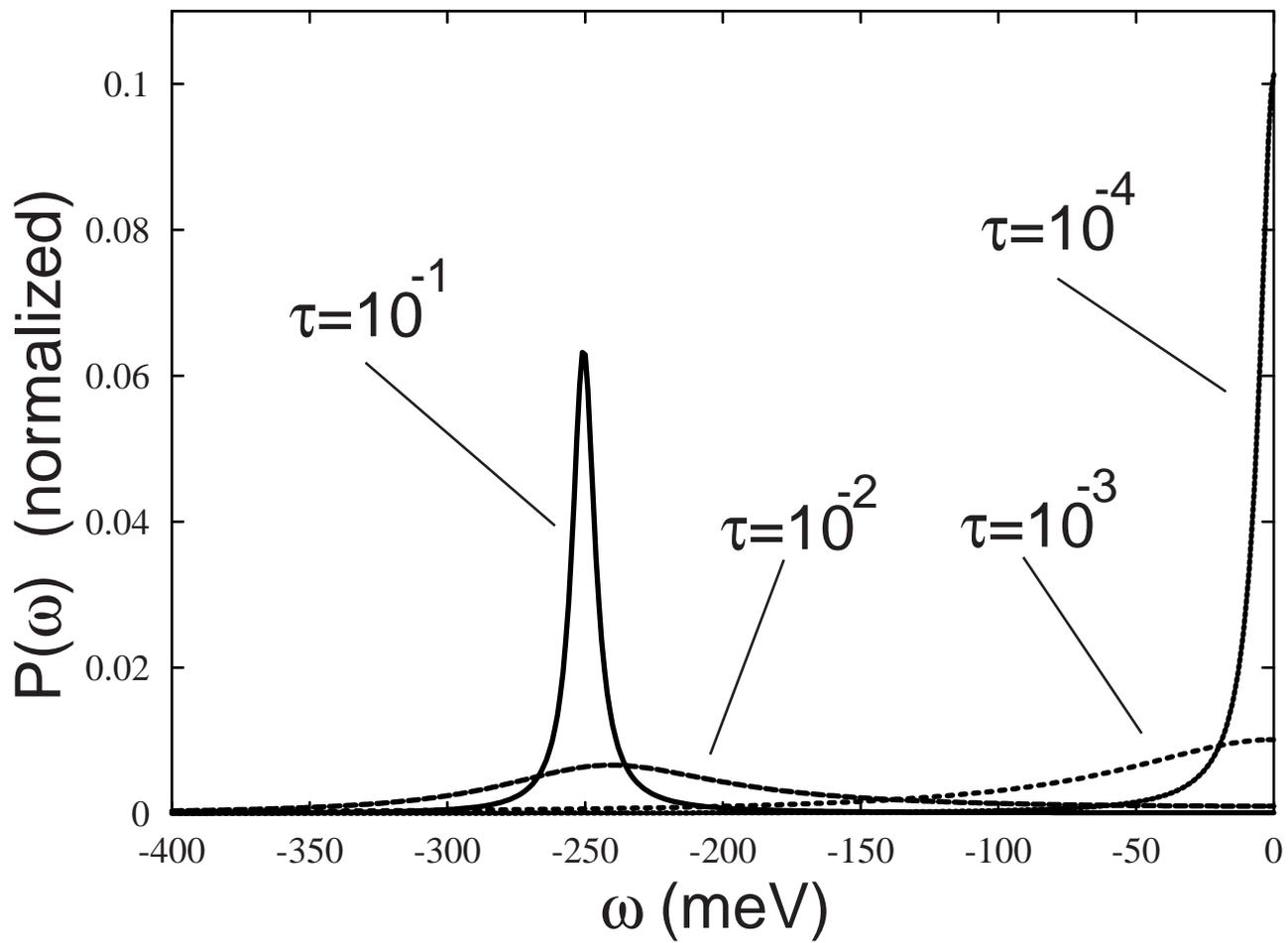
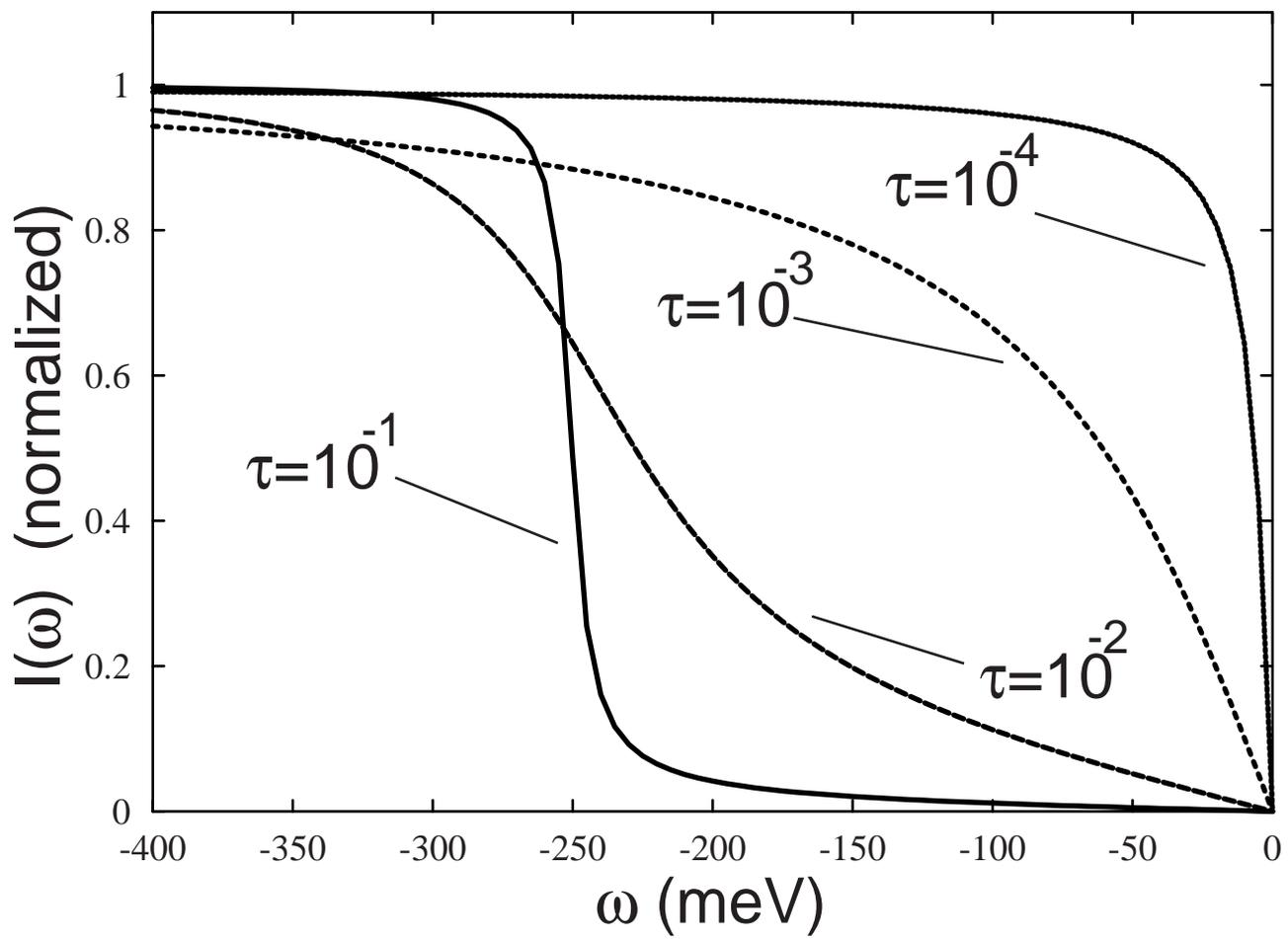
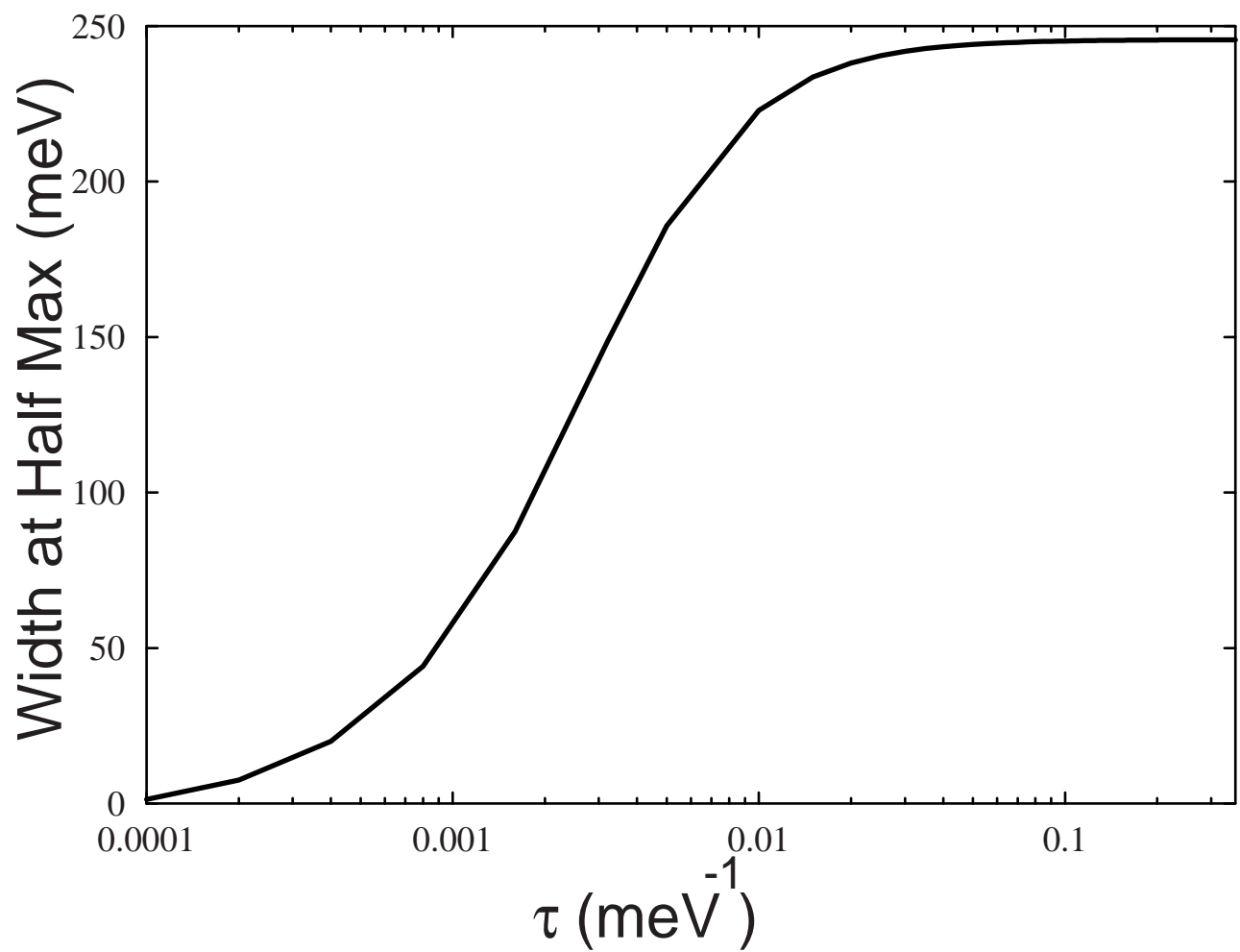
\end{document}